\documentclass[conference]{IEEEtran}
\IEEEoverridecommandlockouts
\usepackage{cite}
\usepackage{amsmath,amssymb,amsfonts}
\usepackage{algorithmic}
\usepackage{graphicx}
\usepackage{textcomp}
\usepackage{xcolor}
\usepackage{multirow}

\newcommand{\noteblue}[1]{\textcolor{black}{{ #1}}}

\def\BibTeX{{\rm B\kern-.05em{\sc i\kern-.025em b}\kern-.08em
    T\kern-.1667em\lower.7ex\hbox{E}\kern-.125emX}}
\begin{document}

\title{
Toward Resilient 5G Networks: Comparative Analysis of Federated and Centralized Learning for RF Jamming Detection
}

\author{\IEEEauthorblockN{Samhita Kuili, Mohammadreza Amini, Burak Kantarci} %Melike Erol-Kantarci}
 \IEEEauthorblockA{School of Electrical Engineering and Computer Science, University of Ottawa, Ottawa, ON, Canada \\
 \texttt{\{skuil016, mamini6, burak.kantarci\}@uottawa.ca}}
 \vspace{-0.35in}
}

\maketitle

\begin{abstract}
Jamming attacks are proliferating and pose a significant threat to the security of 5G and beyond networks. These attacks target 5G radio frequency (RF) domain and can disrupt the communication in wireless networks. While conventional machine learning and deep learning approaches demonstrate its potential for jamming detection, they typically require centralized data collection, compromising the privacy of user equipment (UEs). This work proposes a federated learning (FL)-based jamming detection framework that operates on over-the-air In-phase and Quadrature (IQ) samples extracted from Synchronization Signal Blocks (SSBs) in the RF domain. The framework enables collaborative model training across multiple UEs without sharing raw RF signal data. We adopt Federated Averaging (FedAvg) algorithm to train a 1D convolutional neural network (1DCNN) for effective detection of attacks. Numerical results demonstrate that the proposed FL framework achieves 97\% accuracy and 97\% F1-score, outperforming centralized baselines including MLP, 1DCNN, SVM, and logistic regression, while preserving the data privacy of all participating UEs.

\end{abstract}


\section{Introduction}

With the rapid proliferation of wireless devices under 5G and beyond connectivity, an extensive usage of service plays an essential component across telecommunication infrastructure. The significant technical advancements such as massive multiple-input multiple output (MIMO) \cite{bjornson2020scalable}, millimeter-wave (mmwave) \cite{shen2019miniaturized}, carrier aggregation\cite{goyal2019lte}, software-defined radio \cite{chaudhary2025comprehensive} and non-orthogonal multiple access (NOMA) \cite{sangdeh2020practical} assists in augmenting both quality of service (QoS) and quality of experience (QoE) in wireless networks.  With the increasing reliance on wireless services, security threats have become a critical concern in terms of confidentiality, integrity, and availability of wireless communication. One of the security threats is jamming attacks, which are easy to launch and disrupt the communication channel shared between user equipments (UEs) and 5G base station gNodeB (gNB). These jamming attacks emphasizes the necessity to secure wireless networks from intentional jamming threats. Jamming attacks are predominantly directed at the physical layer (PHY), which overwhelms the legitimate wireless signal by irregular radio jamming signals. These attacks further exploit the inherent vulnerabilities within the Synchronization Signal Blocks (SSBs), which encompass Primary and Secondary Synchronization Signals (PSS and SSS) essential for cell identification and facilitating UE association with the gNB \cite{giordani2018tutorial, pirayesh2022jamming}. Considering the heterogeneity of UEs as well as the privacy and security concerns associated with direct wireless broadcast between UEs and gNB, accurate detection of jamming attacks in radio waveform is vital.

Several conventional machine learning (ML) classifiers, including support vector machine (SVM), \textit{K}-nearest neighbors (KNN), and artificial neural networks (ANN) are exploited for jamming detection \cite{meftah2022federated}. On the contrary, these techniques leverage substantial amount of data for both training and testing. Moreover, as the size of the data increases, the learning and processing time increases significantly, leading to higher computational resource consumption. Furthermore, implementing these algorithms often requires domain expertise to identify the most accurate techniques. Deep learning (DL) showcases enhanced capability in detecting jammed and non-jammed waveform signals \cite{tekbiyik2020real, kulin2018end}. However, DL model demands complex data which makes the overall training process computationally intensive. Additionally, it requires a comprehensive understanding of network topology, training strategies, and parameter selection to effectively utilize appropriate DL frameworks.

Federated learning (FL), a distributed learning approach, enables DL model to train locally without sharing data between UEs. Thus, preserving privacy of participating UEs. In FL, individual UE conduct local model training and share only their learned parameters with a central coordinator.%, thereby preserving data privacy. 
The coordinator aggregates these parameters to form a global model, which is broadcasted back to the UEs. FL performs an iterative training and aggregation process that continues until the global model converges \cite{djaidja2024federated}. Federated average (FedAvg) aggregation algorithm averages the weights of each UE's model during aggregation phase to acquire a robust global model at each training round. This work presents an effective approach for developing an FL-based jamming detection in 5G and beyond networks. %Unlike many studies, the global model is trained over different real-world data sets collected from the 5G and beyond network. 
By leveraging 5G domain knowledge, we exploit SSB, a critical component of the 5G resource grid. This involves processing over-the-air In-phase and Quadrature (IQ) samples of radio waveform and extracting OFDM symbols related to SSB.

The main contributions of the paper are summarized below:

\begin{enumerate}
    \item We introduce a decentralized federated learning jamming detection framework across multiple UEs connected to 5G gNB. The framework operates directly on over-the-air IQ samples extracted from Synchronization Signal Blocks (SSBs) in the 5G RF domain, ensuring data privacy of all participating UEs.
    \item We conduct a comprehensive comparative analysis between the FL framework and centralized AI models to assess the trade-off between privacy preservation and detection capability.
\end{enumerate}

The organization of this paper is as follows.
In Section \ref{RW}, we conduct a review of related work in the field. Section \ref{SM} discusses about the system model adopted for jamming detection. Section \ref{NR} presents the numerical results. In Section \ref{Con} we conclude the article.

\section{Related Work}\label{RW}
Jamming attacks are cyber-attacks that disrupt the wireless communication channel by transmitting noise at the same frequency as the legitimate wireless signals \cite{abou2023privacy}. Additionally, it causes decrease in signal-to-interference-plus-noise ratio (SINR), if the interference or noise is stronger than the signal, therefore resulting into Denial of Service (Dos) of the network. This pose a critical challenge in mission-critical applications. There are different types of jamming types such as constant, deceptive, random, or reactive \cite{meftah2023federated}. Ismail and Reza \cite{ismail2022evaluation} analyze the security performance of three variants of Naive Bayes (NB) in wireless sensor networks (WSNs), while comparing them against artificial intelligence classifiers, including support vector machines (SVM), \textit{K}-nearest neighbors (KNN), and multilayer perceptrons (MLP). The performance of these classifiers is assessed through the metric accuracy and detection probability.  Moreover, Hachimi et al. \cite{hachimi2020multi} propose a multi-stage machine learning intrusion detection system (ML-IDS) in the context of 5G cloud radio access network (C-RAN) environments. This framework identifies and categorizes four jamming attack types by considering WSN-DS dataset. Existing state-of-the-art AI based methods for jamming attacks rely on sensitive raw data which compromise the privacy of the UEs. Furthermore, organization such as network operators are unable to share sensitive data of UEs to enhance and update the AI learning methods due to the widespread risk of adversarial attacks.

%Federated learning (FL) is an advanced paradigm with significant potential to improve machine learning model performance across diverse sectors, primarily through its privacy-preserving collaborative framework. In the domain of wireless network security, FL assists in acquiring a robust jamming attack detection systems without the need to exchange sensitive data among UEs. Consequently, FL allows UEs to collectively contribute to the model development while ensuring privacy preservation. This results in more effective and enhanced intrusion detection capabilities and provides immense security for all UEs.

To address privacy limitations, several studies have explored FL-based approaches for wireless security. Mothukuri et al. \cite{mothukuri2021federated} propose a recurrent neural network (RNN) as a federated learning model to detect anomalies in Internet of Things (IoT) using ModBus network dataset. Moreover, Chen et al. \cite{chen2020intrusion} leverage RNN-based DL models as FL process to ensure high attack detection in wireless edge-enable networks while minimizing the communication cost. Furthermore, Zhang et al. \cite{zhang2021federated} exploit auto-encoder-based models in FL based framework for intrusion detection in the context of Internet of Things (IoT). Additionally, this study leverages NB-IoT dataset for its experiments and has obtained similar performance close to the centralized performance. Djaidja et al. \cite{djaidja2024federated} propose a FL-based intrusion detection system in a 5G and beyond slicing environment by exploiting state-of-the-art FL algorithms, namely FedProx, FedPer, and SCAFFOLD. Additionally, this study takes into account two different scenarios: non-independently and non-identically distributed (Non-IID) and independently and Identically Distributed (IID) due to the diversity of 5G and beyond services. Meftah et al. \cite{meftah2022federated} propose a jamming detection and waveform classification (JDWC) using FL in a distributed tactical wireless networks by focusing on spectral correlation function (SCF) i.e. the frequency domain feature of In-phase and Quadrature signals. Houda et a. \cite{abou2023privacy} propose a jamming resilient intrusion detection model by adopting a secure FL framework while preserving privacy. Additionally, it leverage secure multiparty computation (SMPC) as an aggregation strategy. Furthermore, the framework is evaluated using a wireless sensor networks dataset (WSN-DS) and compared with centralized AI-based methods using metrics F1-score, accuracy, and detection rate.

\noteblue{Over the evolution of 5G security, recent studies demonstrate a shift from conventional data-driven jamming detection to employing standard ontology-based semantic representation and formal verification; for instance, Resource Description Framework (RDF) and Shapes Constraint Language (SHACL)-based threat model frameworks. In addition, such representation assumes 5G entities, interfaces, and security properties that are modeled and validated according to 3GPP specifications {\cite{paskauskas2025decoding, harvanek2024survey, shi2025formal, ko2024toward}} rather than being treated as black-box inputs. However, the aspect of leveraging standard-traceable semantic representations and verification can be considered complementary to our proposed work. While our proposed work emphasizes on PHY-layer jamming detection, where SSB is also a standardized 5G entity, an attack on synchronization signals can be analyzed not only as anomalous RF patterns but also as an indication of severe disruption to explicit standardized functions such as synchronization, cell search access, and identification. Although our proposed method is formulated as a data-driven approach, it can be extensively extended into a standard-traceable knowledge as per 3GPP security reasoning for interpretability. }

While existing studies explore the machine learning and federated learning based methods, several critical gaps remain. Most existing state-of-the-art rely on benchmark datasets such as WSN-DS and NB-IoT, which do not capture the specific characteristics of 5G Radio Frequency signals. These datasets are widely explored for the domain specific task i.e. intrusion detection. However, these works do not consider 5G RF domain, synchronization mechanisms, which are fundamental to UE-gNB association and cell identification. Additionally, \cite{meftah2022federated, meftah2023federated} explore FL-based jamming detection in tactical wireless networks using frequency-domain IQ features, which focuses on spectral correlation functions rather than directly exploiting the 5G resource grid structure. Furthermore, \cite{abou2023privacy} and \cite{djaidja2024federated} address privacy preservation and non-IID data distributions, respectively, but do not evaluate their frameworks on real-world over-the-air 5G signals. Finally, to the best of our knowledge, prior work does not consider privacy-preserving jamming detection using real-world 5G RF data while benchmarking FL against centralized AI baselines under identical experimental conditions. 
%Motivated by these gaps, this work proposes an FL-based jamming detection framework that operates directly on over-the-air IQ samples extracted from SSBs in the 5G RF domain, enabling privacy-preserving detection without reliance on synthetic or non-5G-specific datasets.}

\begin{figure}[t]
\centerline{\includegraphics[scale = 0.43]{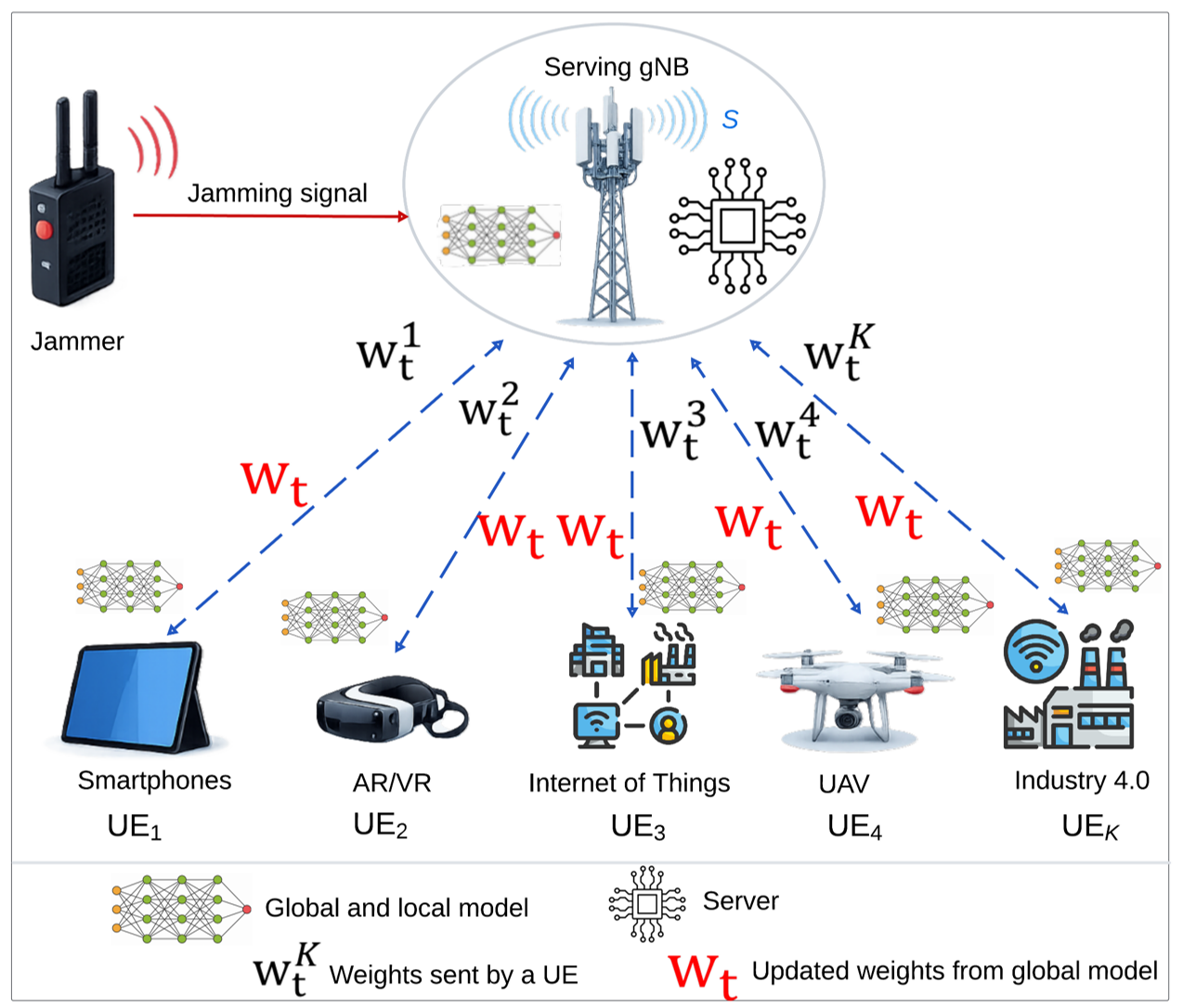}}
\caption{System architecture}
\label{fig1} \vspace{-4mm}
\end{figure}

\section{System Model}\label{SM}

We consider a federated learning (FL) framework deployed in a 5G and beyond wireless network consisting of a server $\mathcal{S}$ employed at gNB and a set of $K$ UEs denoted as $\mathcal{U}=\{UE_1,UE_2,\ldots,UE_K\}$ shown in Fig. \ref{fig1}. These UEs represent heterogeneous devices supporting various 5G and beyond services. \noteblue{Accordingly, the framework can demonstrate a ultra-dense and heterogenous environment of 5G network. } A jammer exists in the system, attempting to impair the legitimate wireless communication between gNB and UEs. To detect any intentional interference or jamming attack, the proposed framework exploits SSB, which is a critical component of UEs for synchronization and cell identification. The objective of the proposed framework is to detect the presence of jamming attacks by analyzing the received SSB observations. The RF data $\mathcal{D}$ can be represented as $\mathcal{D} = (X, Y)$, where $X$ $\in$ $\mathbb{R}^{P \ast Q}$ denotes the size of data with \textit{P} SSB samples and \textit{Q} IQ features and $Y$ $\in$ $\mathbb{R}^{P \ast 1}$ represents the class label indicating whether the received signal is jammed or non-jammed. %Each UE locally processes its observed signals and trains a local learning model using its dataset $\mathcal{D}_k$ .

In 5G and beyond network, each radio cell is characterized by a physical cell identity: cell identity group $N_{ID}^1$ and a identity sector, $N_{ID}^2$. The former can be detected by UE from Secondary Synchronization Signal (SSS), and the later from Primary Synchronization Signal (PSS) respectively. Consequently, the physical cell identity of the serving cell is represented as 
\begin{equation}
N_{ID}^{cell}=3*N_{ID}^1+N_{ID}^2
\end{equation}

This cell identification enables UE to identify and distinguish the serving gNB from neighboring cells and establish synchronization with the transmitter. As jamming can impair the synchronization, identifying SSB is a practical approach to infer if the communication link is under attack.

Let $s(p)$ denote $p^{th}$ IQ sample of the SSB signal transmitted by gNB, which can be represented as

\begin{equation}\label{Eq_tr_wave}
\begin{split}
    s(j)=\sum_{l=0}^{3} s_{l}(m) \quad     j=0,1, \cdots, (l \times m-1) \, 
\end{split}
\end{equation}
\noindent where $s(j)$ denote the discrete-time complex baseband sample corresponding to SSB transmitted by gNB, $s_{l}(m)$ represents $m^{th}$ time domain sample of $l^{th}$ OFDM symbol in SSB, with $m \in \{0,1, \cdots, N_{F\hspace{-.6mm}F\hspace{-.6mm}T} -1\}$ and $N_{F\hspace{-.6mm}F\hspace{-.6mm}T}$ is the size of FFT. Each OFDM symbol $s_l(m)$ contains some data symbols $S_{l,r}$ in the frequency domain which is transformed into time domain as,  \vspace{-4mm}

\begin{equation}\label{Eq_ifft}
\begin{split}
    s_{l}(m)=\frac{1}{N_{F\hspace{-.6mm}F\hspace{-.6mm}T}} \sum_{k=0}^{N_{F\hspace{-.6mm}F\hspace{-.6mm}T}-1} S_{l,k} \,e^{{j2\pi r m}/{N_{F\hspace{-.6mm}F\hspace{-.6mm}T}}} \, 
\end{split}
\end{equation}

The PSS, which is the first OFDM symbol of SSB, i.e. $s_{l}(m) \mid_{l=0}$,  comprises one of three 127-symbol m-sequences and is assigned to the first symbol of each SSB, covering 127 subcarriers. The three potential m-sequences for the PSS are defined as follows \cite{Omri2019}.  \vspace{-4mm}

\begin{equation} \label{Eq_PSS}
    \begin{split}
       S_{l,r+i}\mid_{l=0}=\begin{cases}
       1-2d_p(i) \quad  r \in \{56, \cdots, 182 \}\\
       0      \quad \quad \quad \quad  Otherwise  , 
       \end{cases}
    \end{split}
\end{equation}
\noindent where $d_p(i)$ represents the m-sequences which are given in the 3GPP standard \cite{3gpp.38.211}.

Similar to LTE, 5G SSS serves to detect the physical cell identity. In contrast,  SSS comprises one of 336 127-symbol gold sequences, specifically assigned to the third symbol of each SSB. The 336 potential gold sequences for the SSS are outlined as follows.  \vspace{-4mm}

\begin{equation} \label{Eq_PSS_1}
    \begin{split}
       X_{l,r+i}\mid_{l=3}&=       \big[1-2d_s(i+r_0) mod \, 127\big] \\ 
       & \hspace{5mm}\times \big[ 1-2 d^{\prime }_s (i+r_1) \, mod \, 127  \big] \\
       & \hspace{5mm}  r \in \{56, \cdots, 182 \} \,  ,
    \end{split}
\end{equation}
\noindent where $r_0$ and $r_1$ are derived as,  \vspace{-4mm}

\begin{equation} \label{Eq_PSS2}
    \begin{split}
       r_0=15\Bigg[\frac{N_{ID}^1}{112} \Bigg] +5 N_{ID}^2 \,  ,\\
       r_1=N_{ID}^1 \, mod \, 112 \,  .
    \end{split}
\end{equation}

Furthermore, $d_s(i)$ and $d^{\prime}_s(i)$ can be extracted recursively as stated in 5G standard  \cite{3gpp.38.211}.

At the UE receiver, the observed SSB signal under normal operating condition is expressed as,  \vspace{-4mm}

\begin{equation}\label{Eq_rec_wav}
\begin{split}
    x(j)= s(j) \circledast h(j) + w(j)     
\end{split}
\end{equation}
where $h(j)$ is the channel impulse response and $w(j)$ is the environmental noise. In the presence of a jammer, the observed SSB signal is expressed as,  \vspace{-4mm}

\begin{equation}\label{Eq_rec_wav}
\begin{split}
    x(j)= s(j) \circledast h(j) + w(j) + s_J(j)    
\end{split}
\end{equation} where $s_J(j)$ is the jamming signal.

\begin{figure*}[ht]
    \centering
    \begin{minipage}{.32\textwidth}
        \centering
       \includegraphics[width=1\textwidth]{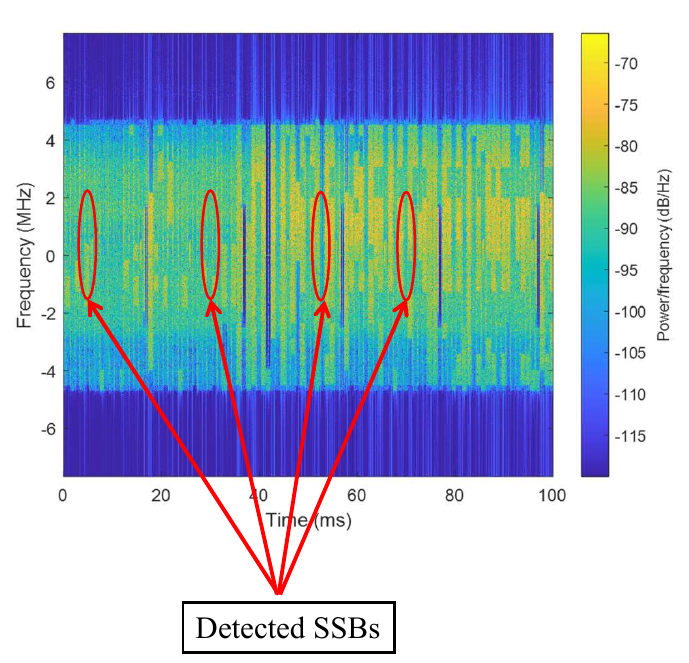}
        \caption{Time-Frequency grid taken from the 5G operator in the absence of jamming signal.}
         \label{fig:SSB_NJ}
    \end{minipage}
    \hfill
    \begin{minipage}{0.32\textwidth}
        \includegraphics[width=\textwidth]{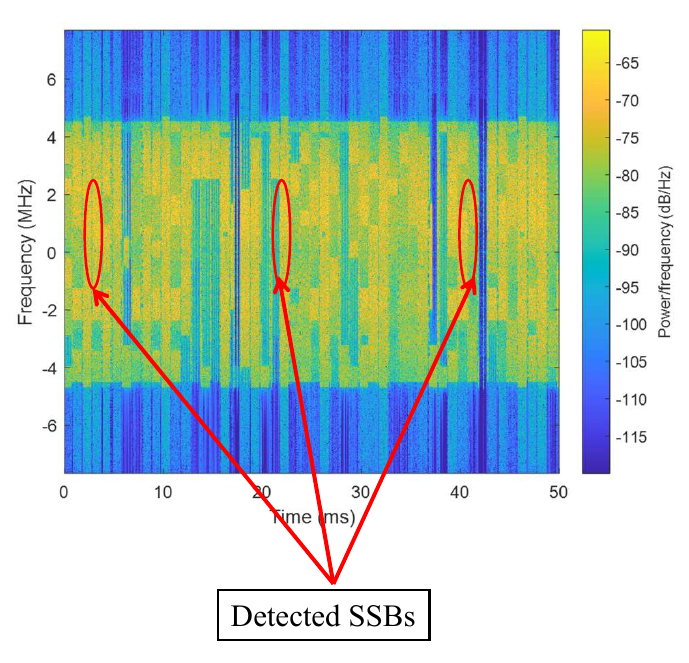}
        \caption{Time-Frequency grid taken from the 5G operator jamming transmit gain is set to -70 dB.}
        \label{fig:SSB_J1.pdf}
    \end{minipage}
    \hfill
    \begin{minipage}{0.32\textwidth}
        \centering
        \includegraphics[width=1\textwidth]{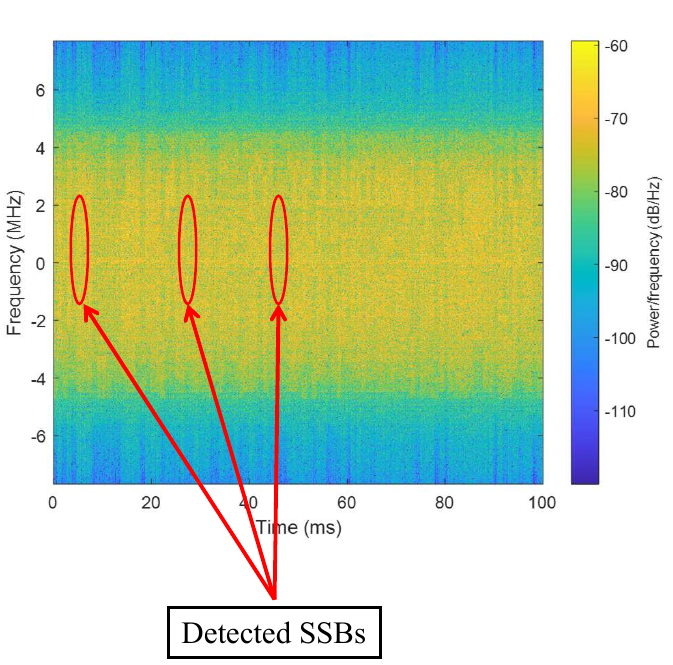}
        \caption{Time-Frequency grid taken from the 5G operator jamming transmit gain is set to -50 dB.}
        \label{fig:SSB_J2}
    \end{minipage} 
\end{figure*}

\begin{figure*}[htp]
    \centering
    \begin{minipage}{.32\textwidth}
        \centering
       \includegraphics[width=1\textwidth]{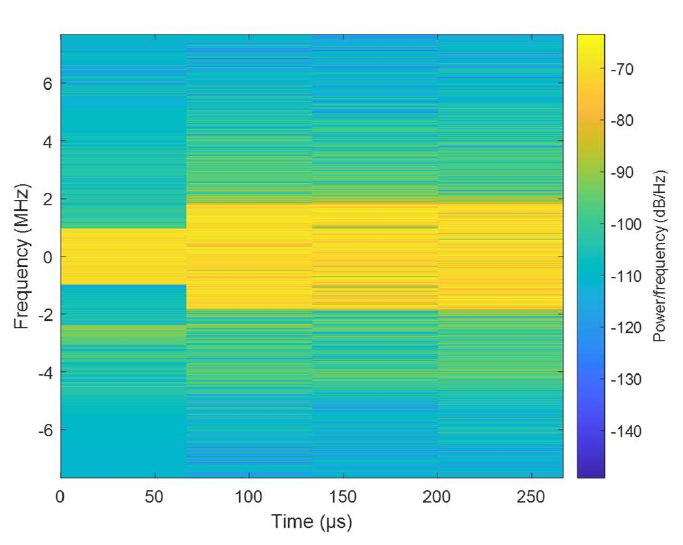}
        \caption{Extracted SSB in the absence of a jamming signal.}
         \label{fig:PSS_NJ}
    \end{minipage}
    \hfill
    \begin{minipage}{0.32\textwidth}
        \includegraphics[width=\textwidth]{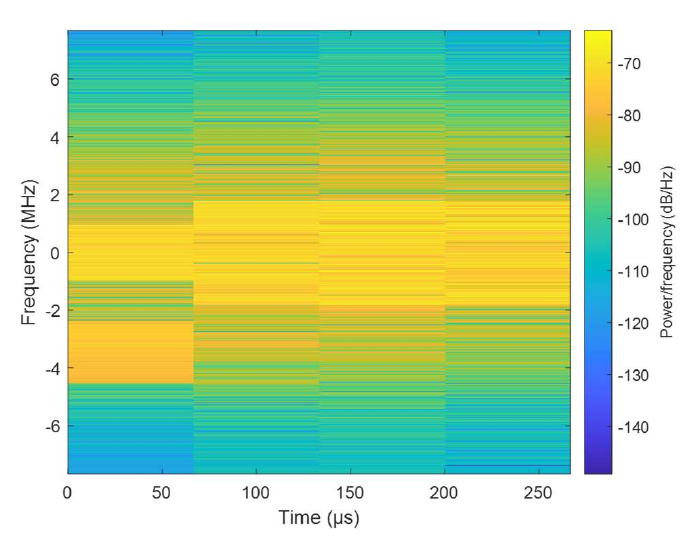}
        \caption{Extracted SSB- jamming transmit gain is set to -70 dB.}
        \label{fig:PSS_J1.pdf}
    \end{minipage}
    \hfill
    \begin{minipage}{0.32\textwidth}
        \centering
        \includegraphics[width=1\textwidth]{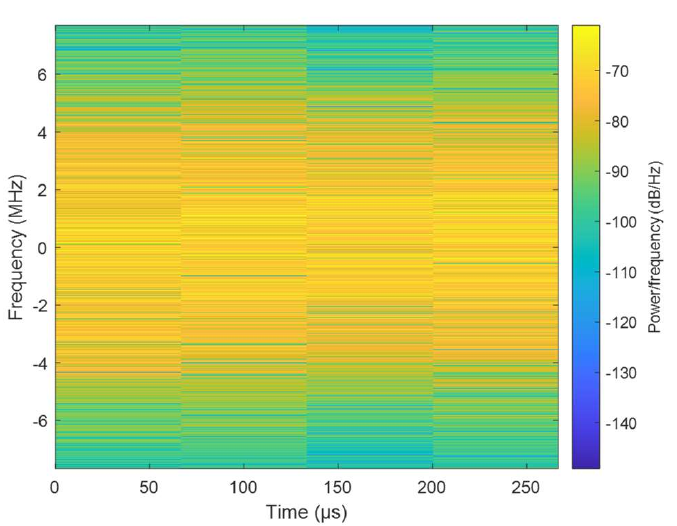}
        \caption{Extracted SSB- jamming transmit gain is set to -50 dB.}
        \label{fig:PSS_J2}
    \end{minipage} 
\end{figure*}

\subsection{Federated learning}

Federated learning enables a single global model to train on data that remains on multiple separate UE in 5G network, often called clients. Each client is responsible to train on local dataset without sharing the raw data to the global model. FL instantiates an optimization problem that aims to minimize the global loss function expressed as a weighted sum of the individual client objectives shown in (\ref{9}).

\begin{equation}\label{9}
\min_{w} f(w) = \sum_{k=1}^{K} \frac{n_k}{n} F_k(w), \quad
F_k(w) = \frac{1}{n_k} \sum_{{\hat{i}} \in P_k} f_{\hat{i}}(w)
\end{equation}

where $K$ is the number of clients, $f_{\hat{i}}(w)=\ell(x_{\hat{i}},y_{\hat{i}};w)$ denotes the loss associated with sample $\hat{i}$, $F_k$ is the objective function of client $k$, $P_k$ is the local dataset of client $k$ with $n_k = |P_k|$, and $n = \sum_{k=1}^{K} n_k$.  Each client \textit{k} computes a local loss over the entire dataset, demonstrating its contribution, which scales to the size of the dataset relative to the total number of samples.

FL aims to acquire prediction performance similar to the centralized algorithms, where all the data are centrally pooled, even though the client distribution typically varies and may not fully represent the global distribution. FL process commences by initializing 1DCNN model as global model at the server. Each client receives the model, trains it locally for a fixed number of iterations $\mathcal{I}$ using its own data, sends the updated parameters (weights) to the server. The global model at the server aggregates these local updates and obtains a new global model, broadcasts it again to clients, and the process repeats for several communication rounds $\mathcal{R}$ until convergence. In this work, we consider the FedAvg aggregation algorithm \cite{mcmahan2017communication}, which averages the weights contributed by different models communicated by clients shown in Fig.\ref{fig1} and (10).

\begin{equation}\label{10}
w_t \leftarrow \frac{1}{n} \sum_{k=1}^{K} n_k w_t^k
\end{equation}

where $w_t $ is the global model at round $t$ obtained by averaging $n_k w_t^k$.

\section{Numerical Results}\label{NR}

\subsection{Experimental Setting}\label{ES}
\subsubsection{Data Generation and RF Signal Visualization}
The data comprises time-frequency spectrograms captured from the 5G band n71, comprising over-the-air signals from network shown in Fig. \ref{fig:SSB_NJ}, Fig. \ref{fig:SSB_J1.pdf}, and Fig. \ref{fig:SSB_J2}. These include both clean signals without interference and signals impaired by Gaussian jamming generated via \textit{ADALM-PLUTO} software-defined radio (SDR) with varying transmission gains to simulate realistic power levels. Samples are obtained using high-resolution spectrum analysis at 15.36 MHz sampling rate. %Fig. \ref{fig:SSB_NJ}, Fig. \ref{fig:SSB_J1.pdf}, and Fig. \ref{fig:SSB_J2} show the 5G Time-Frequency grid taken from the 5G band n71.
Fig. \ref{fig:SSB_NJ} shows 100 ms spectrogram of 5G network under normal operation without jamming interference, demonstrating four SSBs located in the spectrogram. Fig. \ref{fig:SSB_J1.pdf} and \ref{fig:SSB_J2} show another sampled spectrogram for the same network in the presence of the jamming signal, where the jammer transmission gain is set to $-70dB$ and $-50dB$, respectively. Considering that the jamming SDR has a maximum transmission output power of 7dBm and neglecting the cable and the RF combiner loss, the received power at the node will be very low $5 \times 10^{-7} \, mW$ and $5 \times 10^{-5} \, mW$, respectively. Fig. \ref{fig:PSS_NJ}, Fig. \ref{fig:PSS_J1.pdf}, and Fig. \ref{fig:PSS_J2} illustrate the extracted 4 OFDM symbols under three different scenarios: (i) no jamming signal, (ii) jamming with a transmit gain of $-70$ dB, and (iii) jamming with a transmit gain of $-40$ dB. It further highlights that higher jamming power leads to greater degradation of the SSBs, making them increasingly difficult to detect.

\begin{table}[ht]
    \centering
    \caption{Performance comparison of FL over centralized AI-based models}
    %\resizebox{0.5\textwidth}{!}{%
    \begin{tabular}{cccccc}
    \hline
        Models & Class & Precision & Recall & F1-score & Accuracy \\ \hline
        \multirow{2}{*}{MLP} & 0 & 0.92 & 0.99 & 0.951 & \multirow{2}{*}{0.951} \\ 
        ~ & 1 & 0.99 & 0.91 & 0.951 & ~ \\ \hline
        %\multirow{2}{*}{MLP-SVM} & 0 & 0.96 & 1.00 & 0.97 & \multirow{2}{*}{0.974}  \\ 
        %~ & 1 & 0.99 & 0.95 & 0.97 & ~ \\ \hline
        \multirow{2}{*}{1DCNN} & 0 & 0.93 & 1 & 0.96 & \multirow{2}{*}{0.959} \\ 
        ~ & 1 & 1 & 0.92 & 0.96 & ~ \\ \hline
        %\multirow{2}{*}{RF} & 0 & 1 & 1 & 1 & \multirow{2}{*}{0.997} \\ 
        %~ & 1 & 1 & 1 & 1 & ~ \\ \hline
        \multirow{2}{*}{SVM} & 0 & 0.86 & 1 & 0.92 & \multirow{2}{*}{0.917} \\ 
        ~ & 1 & 0.99 & 0.84 & 0.91 & ~ \\ \hline
        \multirow{2}{*}{LR} & 0 & 0.84 & 0.99 & 0.91 & \multirow{2}{*}{0.906} \\ 
        ~ & 1 & 0.99 & 0.82 & 0.9 & ~ \\\hline
        \multirow{2}{*}{\textbf{FL}} & \textbf{0} & \textbf{0.94} & \textbf{1} & \textbf{0.97} & \multirow{2}{*}{\textbf{0.97}} \\ 
        ~ & \textbf{1}& \textbf{1} & \textbf{0.94} & \textbf{0.97} & ~ \\\hline
    \end{tabular}
    %}
    \label{tab2} 
\end{table}

\subsubsection{FL setup}The simulation is considered by assuming $K = 9$ clients (UEs) for instantiating FL process for jamming detection. The real-world RF data comprises $P$ = 14129 SSB samples and  $Q$ = 54 IQ features. We assume a train set of 11303 samples and test set of 2826 samples under 80:20 split ratio. The train data is partitioned uniformly across clients. Each client comprises local dataset with SSB observations and information of class labels jammed (1) and non-jammed (0) signals. We conducted model training and experiments using NVIDIA RTX A4000, TensorFlow's GPU-accelerated computations through CUDA 11.2 and cuDNN 8.1.

\subsection{Experimental Results} 
\begin{figure}[t]
\centerline{\includegraphics[scale = 0.28]{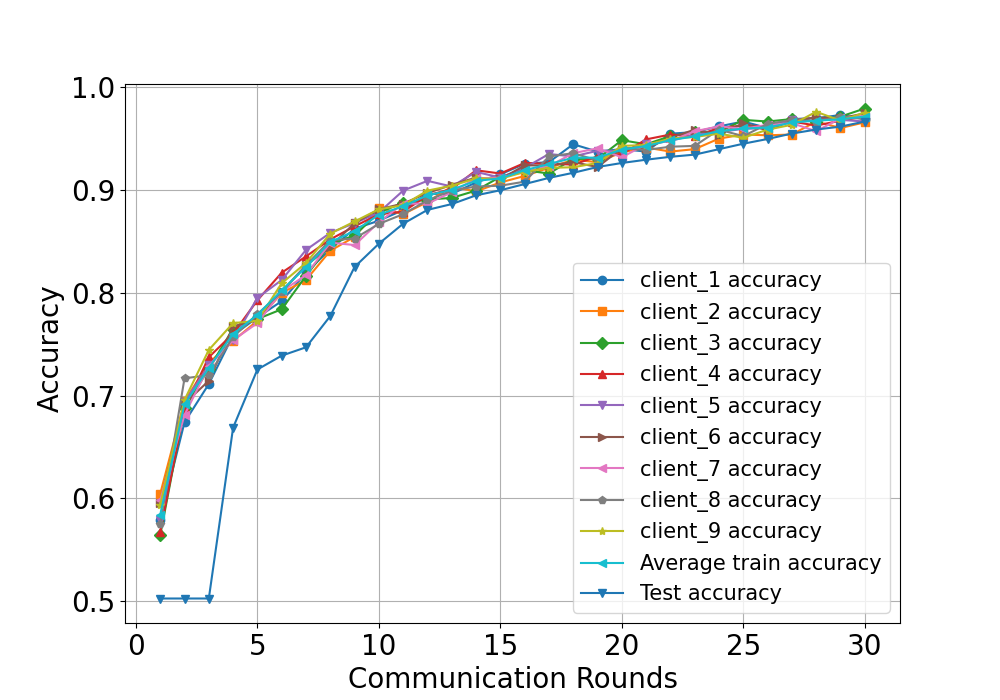}}
\caption{Accuracy of FL with communication rounds}
\label{fig7} \vspace{-4mm}
\end{figure}

\begin{figure}[t]
\centerline{\includegraphics[scale = 0.28]{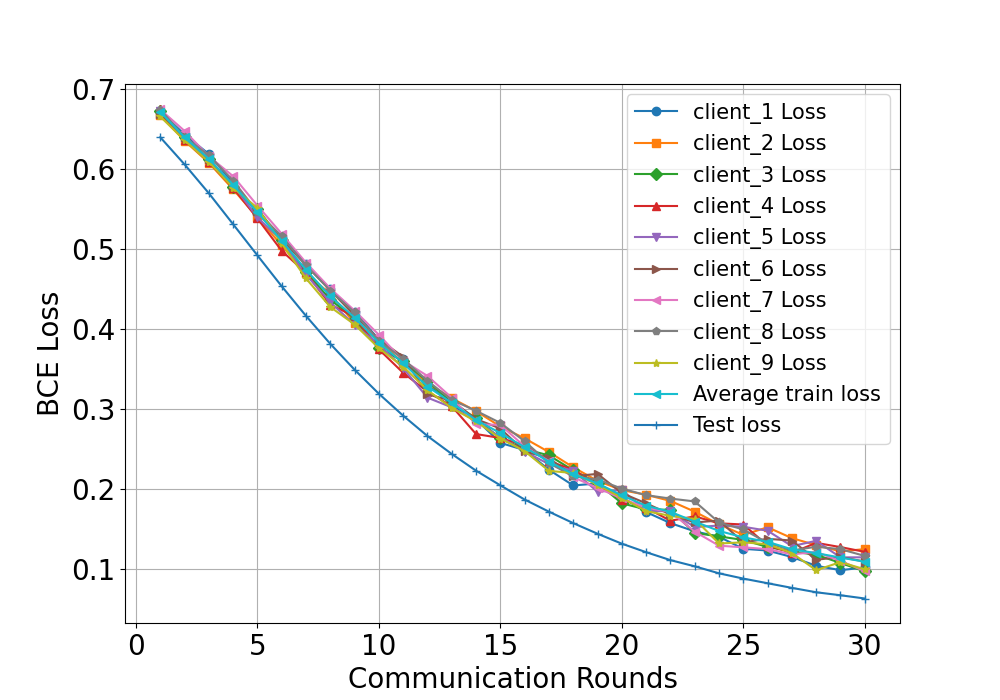}}
\caption{Loss of FL with communication rounds}
\label{fig8} 
\end{figure}

\begin{table}[t]
\caption{Parameters/Hyperparameters FL}
\label{tab1}
\centering
%\resizebox{5.5cm}{!}{
\begin{tabular}{c|c}
\hline
\textbf{Parameter/ Hyperparameter} & \textbf{Value/Setting} \\ \hline

Batch Size              & 16                    \\ 
Optimizer         & SGD                   \\ 
Learning Rate     & 0.001                 \\ 
%\textbf{Server Optimizer}         & SGD                   \\ 
%\textbf{Server Learning Rate}     & 1.0                   \\ 
Loss Function            & Binary Cross Entropy  \\ 
Number of local iterations $\mathcal{L}$ & 1 \\
\hline

\end{tabular}
%}\vspace{-0.3cm}\
\end{table}

\subsubsection{Performance of FL}
In Fig. \ref{fig7}, FL shows the training performance of local models (1DCNN) and evaluation of global model (1DCNN) over $\mathcal{R}$ = 30 communication rounds with the help of metric accuracy. The local training of FL follows the parameters chosen as discussed in Table \ref{tab1}. The average training accuracy for all clients increases from 58.40\% to 97.16\%. Additionally, the global test accuracy increases from 50.24\% to 96.67\%. Moreover, the performance of FL is demonstrated with the help of binary cross entropy loss. Fig. \ref{fig8} showcases average train loss decreases from 0.671 to 0.109 for all clients. Additionally, the global test loss reduces from 0.639 to 0.063.

\subsubsection{Performance comparison of FL with centralized AI-based models}

In this section, we compare the performance of FL with centralized AI-based schemes in detecting jamming attacks. We utilize various metrics, such as accuracy, precision, recall, and F1-score as shown in Table \ref{tab2}. The centralized AI-based models, including MLP \cite{almomani2016wsn}, 1DCNN \cite{liu2019deep}, SVM and LR \cite{killeen2025iot} are considered for comparative analysis. These centralized models compromise data privacy as they are trained on raw SSB observations obtained collectively from all clients.
FL achieves F1-score of 97\%, and accuracy of 97\%, outperforming the performance of centralized models. FL showcases close and enhanced performance of jamming detection as compared to centralized models in 5G and beyond networks.

\section{Conclusion}\label{Con}
In this article, we presented a federated learning framework for jamming attack detection in 5G and beyond wireless networks. We have evaluated our proposed framework on 5G Radio Frequency signal by exploiting a 5G resource grid to detect jamming attacks. Additionally, we have employed FedAvg aggregation algorithm to train 1DCNN model between UEs and server, achieving an accuracy and F1-score of 97\%. Furthermore, we have compared with centralized AI models, including MLP, 1DCNN, SVM, and logistic regression. The comparative analysis have shown that federated learning can achieve close and enhanced detection performance as compared to centralized models while preserving privacy of each UE participating in FL process. These findings underscore the suitability of FL-based frameworks for security-critical applications in next-generation wireless networks. Future work will further explore \noteblue{interpretability of IQ-features and SSB observations using Knowledge Graph,} robustness under non-IID data distributions, adversarial FL attacks such as data and model poisoning, and the integration of more advanced aggregation strategies to further strengthen the framework under realistic heterogeneous network conditions.

\section*{Acknowledgment}
This work was supported in part by the Natural Sciences and Engineering Research Council of Canada (NSERC) under Discovery and CREATE TRAVERSAL programs.

\bibliographystyle{IEEEtran}
%\bibliography{ref}

\end{document}